\documentclass[12pt,dvips]{article}

\usepackage{amsmath,amssymb,exscale}
\usepackage{array,multicol}
\usepackage{afterpage,float,flafter}
\usepackage{epsfig,rotating,pifont}
\usepackage{cite}
\@addtoreset{equation}{section} \makeatother

\title{
\vspace{1cm}
\Large\textbf{Spin and a Running Radius in RS1}
\vspace*{.5cm}
\author{\large \textbf{
Adam Lewandowski\footnote{email: adaml@pha.jhu.edu}
\mbox{  }and Michele Redi\footnote{email: redi@pha.jhu.edu}}\\
\emph{
Department of Physics and Astronomy} \\
\emph{Johns Hopkins University} \\
\emph{3400 North Charles St}. \\
\emph{Baltimore, MD 21218-2686}}}

\date{}
\begin{document}
\maketitle
\thispagestyle{empty}
\vspace*{.5cm}

\begin{abstract}
We develop a renormalization group formalism for the compactified
Randall-Sundrum scenario wherein the extra-dimensional radius
serves as the scaling parameter. Couplings on the hidden brane
scale as we move within local effective field theories with
varying size of the warped extra dimension.  We consider this RG
approach applied to $U(1)$ gauge theories and gravity. We use this
method to derive a low energy effective theory.
\end{abstract}

\newpage
\renewcommand{\thepage}{\arabic{page}}
\setcounter{page}{1}

\section{Introduction}

The compactified Randall-Sundrum (RS) scenario provides a natural
explanation for the separation of the weak/Planck scales observed
in nature \cite{rs1}.  The warped space-time presents a
difficulty, however, in evaluating Feynman diagrams and estimating
their size. In particular, the exponentially small warp factor
responsible for setting the large hierarchy can appear in ways
that are difficult to predict, having dramatic consequences for
considerations such as the stability of the RS hierarchy and gauge
unification \cite{adshard}. Recently, a novel approach was begun
for studying this scenario wherein holographic renormalization
group (RG) ideas \cite{holorg,skenderis} were used to identify
scaling effects and simplify power counting in an effective field
theory approach \cite{runrs}. Here we continue this program,
extending the analysis to RS theories with bulk photons and
gravitons.

We will consider the geometry of the RS1 scenario where five
dimensional AdS space is compactified on an orbifold
$S_1/\mathbb{Z}_2$ with two branes (the hidden and the visible
brane) located at the fixed points. Following Ref. \cite{runrs},
we match to an effective theory with a smaller proper distance
between the visible brane and an effective hidden brane. This
matching is accomplished by requiring that the classical solutions
of the field equations are identical in the overlapping region.
This procedure requires an effective hidden brane action local in
the fields and their derivatives and gauge invariant (or general
covariant in the case of spin 2 fields). The couplings on the
effective hidden brane flow at the classical level in the sense of
a Wilsonian RG flow. In fact, our method closely resembles
Wilson's picture of renormalization, integrating out degrees of
freedom close to the hidden brane we flow in the space of local
theories with various radii. In running the radius down to a
negligible extra-dimensional size, the large warp factor is
integrated into the hidden brane couplings in predictable ways
simplifying power counting analysis. We may identify in the
running of these couplings certain features of the AdS/CFT duality
\cite{adscft} applied to the RS scenario \cite{rscft}. In fact, by
moving the visible brane, the same method can be applied to
holographic computations in the RS/CFT correspondence, as shown in
\cite{redi} for the case of scalar fields.

The analysis of photons and gravitons is similar to the analysis
of bulk scalar fields of Ref.~\cite{runrs} in many respects.
Particular to the case of the massive scalar however, is the
existence of an attractive local fixed point. We will see that
such a fixed point is absent in gauge theories or any theory
containing massless fields. From the CFT side the absence of a
fixed point is understood as the dual of these RS theories
contains a massless field breaking conformal invariance at any
scale. In the brane running approach developed here, we find that
it is precisely the necessity of the massless zero mode in gauge
theories that precludes the existence of a fixed point.

We find an equivalent field theory description of these bulk gauge
and gravity theories with an attractive local fixed point. By
adding a massless hidden brane field to the bulk photon or
graviton theories we treat the zero mode explicitly in the
running. This massless field decouples from the massive
Kaluza-Klein (KK) tower at the fixed point. One can use the
attractive nature of the fixed point to simplify the running to
find sensible low energy effective theories.

The paper is organized as follows. In Section \ref{photon} we
discuss the matching procedure for a free bulk photon propagating
in the RS1 geometry. We consider spin 2 fields in the linearized
approximation in Section \ref{graviton}. In Section
\ref{improvement} we introduce a method of improving the effective
field theory description by adding a massless hidden brane field.
We examine this method for the case of massless scalar fields and
$U(1)$ gauge theories. Section \ref{conclusions} contains a brief
discussion of the results. In the Appendix we extend the procedure
to the case of scalar QED with general hidden brane interactions.


\section{The free photon}
\label{photon}

We consider a fixed AdS$_5$ background with metric
\begin{equation}
ds^2 = \frac{1}{(kz)^2} ( \eta_{\mu \nu} dx^{\mu} dx^{\nu} -
dz^2). \label{metric}
\end{equation}
The $z$ coordinate parameterizes the direction in the orbifolded
space. The space is bounded by a hidden brane at $z=z_h =1/k \sim
{\mathcal O}(1/M_{Plank})$ and a visible brane at $z=z_v \sim
{\mathcal O}(1/{\rm TeV})$.

We examine the case of a free $U(1)$ gauge field with action
\begin{equation}
S=S_{bulk}+S_{hid}+S_{vis},
\end{equation}
where
\begin{eqnarray}
S_{bulk}&=&-\frac {1} {4 g_5^2} \int d^4x \int_{z_v}^{z_h} dz
\sqrt{G} G^{M N}G^{S P} F_{M S} F_{N P},\\
S_{hid}&=&-\frac {1} {4 g_5^2 k} \int d^4x \sqrt{-g_h}g_h^{\mu
\nu} g_h^{\sigma \delta} F_{\mu \sigma} \tau_h F_{\nu \delta}, \\
S_{vis}&=&-\frac {1} {4 g_5^2 k} \int d^4x \sqrt{-g_{v}} g_v^{\mu
\nu} g_v^{\sigma \delta} F_{\mu \sigma} \tau_v F_{\nu \delta},
\label{photonaction}
\end{eqnarray}
and
\begin{eqnarray}
g^h_{\mu \nu}(x) & = & G_{\mu \nu}(x,z=z_h), \\
g^v_{\mu \nu}(x) & = & G_{\mu \nu}(x,z=z_v).
\end{eqnarray}
The brane actions are the most general quadratic actions
compatible with gauge invariance and $4D$ general covariance. The
coefficient $\tau_h$ is a series of derivative operators,
\begin{equation}
\tau_h = \tau_h^{(0)} - k^{-2} \tau_h^{(2)} g_h^{\mu \nu}
\partial_{\mu} \partial_{\nu} + k^{-4} \tau_h^{(4)} g_h^{\mu \nu}
g_h^{\rho \sigma} \partial_{\mu}
\partial_{\nu} \partial_{\rho} \partial_{\sigma}+...
\label{tauh}
\end{equation}
and similarly for $\tau_v$. We assume $A_\mu$ to be even and $A_z$
odd on the orbifold.

The equations of motion are given by
\begin{eqnarray}
\partial_M(\sqrt{G}G^{M N} G^{5 P} F_{N P}) &=& 0 \\
\partial_M(\sqrt{G}G^{M N} G^{\delta P} F_{N P})&=&-\frac 1 {k}
(\tau_h \delta(z-z_h)+\tau_v \delta(z-z_v))
\partial_\mu(\sqrt{-g_h} g^{\mu \nu} g^{\delta \sigma} F_{\nu
\sigma}).
\label{photonequation}
\end{eqnarray}
These equations greatly simplify in the standard gauge
$A_z(x,z)=0$ and $\partial^\mu A_\mu(x,z)=0$. This is a consistent
gauge choice as there is no Aharonov-Bohm phase around the fifth
dimension for $A_z$. It is the most
convenient choice for our considerations here.

Fourier transforming the 4D Minkowskian directions, equation
(\ref{photonequation}) becomes the boundary independent
differential equation
\begin{equation}
(z^2 \partial_z^2-z\partial_z+q^2z^2)A_\mu(q,z)=0,
\end{equation}
with the following boundary conditions determined by the brane
actions
\begin{eqnarray}
z \partial_z A_{\mu}(z)\Big|_{z=z_h} &=& -\frac {(qz_h)^2} {2}
\tau_h(q^2 z_h^2)
A_{\mu}(z_h) \\
z\partial_z A_{\mu} (z)\Big|_{z=z_v} &=& +\frac {(qz_v)^2} 2
\tau_v (q^2 z_v^2) A_{\mu}(z_v).
\end{eqnarray}
where $q$ is the $4D$ momentum.

We now consider an effective theory with a new hidden brane
bounding the space at $a>z_h$. We require the new theory to
reproduce the same physics from the point of view of an observer
living at $z>a$. We take the same visible brane and bulk action as
in (\ref{photonaction}) and introduce the following effective
hidden brane action,
\begin{equation}
S_a=-\frac {1} {4 g_5^2 k} \int d^4x \sqrt{-g_a}g_a^{\mu \nu}
g_a^{\sigma \delta} F_{\mu \sigma} \tau(a) F_{\nu \delta},
\end{equation}
where $g_a$ is the induced metric at $z=a$. The coupling $\tau(a)$
has a derivative expansion similar to (\ref{tauh}),
\begin{equation}
\tau(a)=\sum_{j} (-1)^j \tau^{(2j)}(a) (k^{-2} g_{a}^{\mu \nu}
\partial_{\mu} \partial_{\nu})^j.
\label{tau}
\end{equation}
Notice that $\tau(a)$ contains an explicit dependence on $a$,
coming from the warped metric and an intrinsic one from the
individual couplings $\tau^{(2i)}(a)$.

The effective brane action implies the boundary condition at $a$
on the fields,
\begin{equation}
z \partial_z A_{\mu}(z) \Big|_{z=a} = -\frac {(qa)^2} {2} \tau(q^2
a^2, a) A_{\mu}(a). \label{braneaction}
\end{equation}
By assumption the solutions of the classical equations of motion
in the restricted region $z\in (a,z_v]$ are unchanged under the
motion of the brane. This condition gives a \emph{classical} flow
equation for the coupling $\tau$ similar to a $4D$ RG flow.

Explicitly, we take the logarithmic derivative with respect to $a$
of the boundary condition (\ref{braneaction}). Using the bulk
equation of motion to eliminate the second derivatives of the
fields one finds
\begin{equation}
a \frac {d} {da} \tau(q^2a^2,a)=\frac {(qa )^2} {2} \tau^2+2.
\label{gaugeflow}
\end{equation}
This equation can be interpreted as a set of coupled flow
equations for the individual couplings $\tau^{(2j)}(a)$ in
(\ref{tau}), obtained by expanding in powers of $q$
(\ref{gaugeflow}),
\begin{equation}
a \partial_a \tau^{(2j)} = -2j \tau^{(2j)}+\frac 1 2 \sum_i
\tau^{(2 i)} \tau^{(2j-2i-2)}+2 \delta_{j,0}.
\label{couplingsflow}
\end{equation}
The flow equations for the couplings $\tau^{(2j)}$ have a lower
diagonal form. Operators mix with operators of lesser or equal
dimension. This feature will be preserved in the interacting
theory considered in detail in the Appendix.

We may search for fixed points to this flow satisfying
\begin{equation}
\partial_a \tau^{(2j)}(a)=0.
\label{gaugefp}
\end{equation}
It is not hard to check that
\begin{equation}
\tau^*=-2\frac {J_0(qa)} {qaJ_1(qa)}.
\end{equation}
is a solution to (\ref{gaugefp}). There are also other solutions
to (\ref{gaugefp}) with different linear combinations of Bessel
functions with the same index but these have an expansion non
polynomial in momentum. These correspond to non-local effective
brane operators and so, in the spirit of the local effective field
theory, they are rejected. They are however important in the
context of the AdS/CFT correspondence \cite{redi}.

The first term in the expansion of $\tau^*$ is proportional to
$1/q^2$ corresponding to a brane localized mass term. A gauge
theory can only approach such a fixed point if the gauge symmetry
is broken. In fact, if we add a mass term for the gauge field on
the hidden brane, that is $\tau^{(-2)}(z_h)\ne 0$, then $\tau(a)$
approaches $\tau^*$ for large $a$. If the gauge symmetry is
unbroken however there is no fixed point to the flow.

We can integrate analytically equation (\ref{gaugeflow}),
\begin{equation}
\tau(a)=-2 \frac {Y_0(q a)+ B J_0(q a)} {qa(Y_1(qa)+B J_1(qa))}.
\label{tausolution}
\end{equation}
The boundary condition at the hidden brane
\begin{equation}
\tau(q^2a^2,a)|_{a=z_h} = \tau_h,
\end{equation}
determines the integration constant $B$,
\begin{equation}
B=- \frac {Y_0(q z_h)+q z_h \tau_h Y_1(q z_h)} {J_0(q z_h)+q z_h
\tau_h J_1(q z_h)}.
\end{equation}
Running the hidden brane to the visible brane the entire
contribution to the action is given by the brane actions since the
bulk contribution becomes negligible. Notice that the zeros in
$q^2$ of $\tau(a)|_{a=z_v}+\tau_{z_v}$ are the $4D$ masses of the
KK tower of the gauge field. This follows from the fact that the
exact propagator in the effective field theory coincides with the
brane-to-brane propagator of the full theory.

The leading term in the momentum expansion of (\ref{tausolution})
is given by
\begin{equation}
\tau^{(0)}(a)=2 \log\Big(\frac {a} {z_h}\Big)+ \tau^{(0)}_h.
\label{logrunning}
\end{equation}
This can be easily obtained from (\ref{couplingsflow}). Running to
$a=z_v$ we find that the $4D$ effective gauge coupling is:
\begin{equation}
\frac 1 {g_4^2}=\frac {2 \log\big(\frac {z_v}
{z_h}\big)+\tau_h^{(0)}+\tau_v^{(0)} } {g_5^2 k}.
\end{equation}
which reproduces the zero mode calculation. The kinetic terms on
the hidden or on the visible brane modify the effective $4D$ gauge
coupling by additive corrections. As expected the coupling
$\tau(a)$ has a local expansion in momentum.

The solution of equation (\ref{tausolution}) gives the exact flow
for any $\tau_h$.  It is useful to examine the flow equations in
the form of equation (\ref{couplingsflow}) to estimate the size of
the coefficients $\tau^{(2j)}(a)$.  We see first that $\tau^{(0)}
\sim \log a$.  Iterating in equation (\ref{couplingsflow}) we
deduce that $\tau^{(2j)} \sim (\log a)^{j+1}$ and no larger.  In
terms of external momentum $q$ the operator with coefficient
$\tau^{(2j)}$ will power count as $(q^2 a^2 \log (a/z_h))^{j+1}$.
The coefficients are enhanced by large logs.  However, the
effective theory is valid in the regime $q^2 a^2 \log (a/z_h) <
1$. In the Appendix we show that this conclusion holds in the
interacting theory as well.

It is important to understand the physical origin of the
logarithmic enhancement of the $q^2 a^2$ expansion. This is a
result of the fact that the massless zero mode of a gauge field is
not localized near the visible brane but flat throughout the
extra-dimensional space. The KK modes of a gauge field couple to
the visible brane parametrically stronger than the zero mode
\cite{Rizzo1}. As a consequence, they become important at energies
below the mass of the first KK mode. The existence of a
non-localized zero mode increases the sensitivity to the hidden
brane position and decreases the region of validity of the
effective theory lowering the cutoff to $1/(a \log(a/z_h))$. A
similar story occurs whenever we consider theories with massless
modes. In Section \ref{improvement} we will show how to find
improved effective theories with massless zero modes having a
cutoff at the scale $1/a$ (no $\log$ suppression).

\section{Gravity}
\label{graviton}

The same RG formalism can also be applied to gravity. Here we
consider the linearized fluctuations of the gravitational field.

We start by briefly reviewing the Randall-Sundrum solution of
Einstein's equations. Following the conventions of \cite{rs1} we
take the gravitational action to be
\begin{equation}
S=\int d^4x dz \sqrt{G} (2 M_5^3 R-\Lambda)-T_h \int d^4x
\sqrt{-g_h}-T_v \int d^4x \sqrt{-g_v},
\label{gravityaction}
\end{equation}
where $T_h$ and $T_v$ are the tensions of the hidden and visible
brane, necessary to reproduce the singularities of the metric at
the fixed points of the orbifold. The metric
\begin{equation}
ds^2=\frac 1 {(kz)^2}(\eta_{\mu \nu} dx^\mu dx^\nu-dz^2),
\label{ansatz}
\end{equation}
defined on the orbifold is a solution of Einstein's equations when
the brane tensions are tuned as
\begin{equation}
T_h=-T_v=24 M_5^3 k
\end{equation}
and the bulk cosmological constant is
\begin{equation}
\Lambda=-24M_5^3 k^2.
\end{equation}
The zero mode fluctuations of the metric (obtained by replacing
$\eta_{\mu \nu}$ with $g_{\mu\nu}(x)$ in (\ref{ansatz})) reproduce
ordinary four dimensional gravity with $4D$ Planck mass given by
\begin{equation}
M_4^2=\frac {M_5^3} {k^3}\Big(\frac 1 {z_h^2}-\frac 1
{z_v^2}\Big). \label{planckmass}
\end{equation}

We now consider the linearized tensor fluctuations of the metric
$G_{\mu\nu}(x,z)=1/(kz)^2~\eta_{\mu\nu}+h_{\mu\nu}(x,z)$. Choosing
the gauge $\partial^\mu h_{\mu \nu}=0$ and $h^{\mu}_{\mu}=0$, the
equations of motion assume a particularly simple form,
\begin{equation}
\Big(\frac 1 2 (z^2\partial_z^2+z\partial_z-4-z^2 \square)+2
z\delta(z-z_h)-2 z \delta(z-z_v)\Big) h_{\mu \nu}(x,z)=0
\label{lingravity}
\end{equation}
where $q$ is the four dimensional momentum. The delta-Dirac terms
include the contributions coming from the singularities of the
derivatives of the background metric as well as the brane terms.
This is different from the case of scalar and gauge fields where
delta-Dirac terms in the equations of motion arise solely from
brane contributions.

The effective theory with a smaller radius has an effective hidden
brane located at $a$ where $z_h<a<z_v$. The quadratic hidden brane
action at $a$ is given by
\begin{equation}
S_a=2 M_5^3 k \int \sqrt{-g_a}  h_{\mu \nu} (3\alpha-4){\cal
O}^{\mu\nu\rho\sigma} h_{\rho \sigma}. \label{gravitybrane}
\end{equation}
Here we have introduced the operator ${\cal O}^{\mu\nu\rho\sigma}$
in order to preserve the invariance of the action under
infinitesimal diffeomorphisms. In the gauge $\partial^\mu h_{\mu
\nu}=0$ and $h^{\mu}_{\mu}=0$ considered above, ${\cal
O}^{\mu\nu\rho\sigma}$ simply becomes
$g_a^{\mu\rho}g_a^{\nu\sigma}$. This brane action is derived
generalizing the hidden brane tension to a running coupling $T(a)
= 24 M^3 k \alpha(a)$, where $\alpha(a)$ represents the series of
derivative couplings,
\begin{equation}
\alpha(a)=\sum_j (-1)^j
\alpha^{(2j)}(a)(k^{-2}g_a^{\mu\nu}\partial_\mu\partial_\nu)^j.
\end{equation}
These couplings are the coefficients of some linear combination of
higher derivative general covariant operators.

The gauge fixed linearized equations of motion derived from the
effective theory are
\begin{equation}
\Big(\frac 1 2 (z^2
\partial_z^2+z\partial_z-4-z^2 \square)-(6\alpha-8) z \delta(z-a) - 2 z \delta(z-z_v)\Big) h_{\mu
\nu}(x,z)=0. \label{quadraticaction}
\end{equation}
We obtain the boundary condition at $a$
\begin{equation}
z\partial_z h_{\mu\nu}(z)\Big|_{z=a}=(6 \alpha -8)h_{\mu\nu}(a).
\end{equation}
Taking the logarithmic derivative with respect to $a$ of this
equation and using the bulk equations of motion we find the flow
for $\alpha$
\begin{equation}
a \frac {d} {da} \alpha(q^2 a^2,a)=-6 \alpha^2 + 16 \alpha - 10 -
\frac {1} {6} (q a)^2
\label{gravityflow}
\end{equation}
This flow has a local fixed point given by
\begin{equation}
\alpha^*=1+\frac 1 6 qa \frac {J_1(qa)} {J_2(qa)} = \frac 5 3
-\frac {(qa)^2} {36}+ \dots. \label{gravityfp}
\end{equation}
The value of the brane tension at the fixed point $\alpha^{(0)*}$
is inconsistent with the Poincar\'e ansatz which requires
$\alpha^{(0)}(z_h)=1$. In fact, there is no local fixed point
consistent with $4D$ Poincar\'e invariance.

The solution of equation (\ref{gravityflow}) is
\begin{equation}
\alpha(a)=\frac {qaY_1(qa)+6Y_2(qa)+C(6J_2(q a)+qa J_1(q a))}
{6Y_2(q a)+6 C J_2(q a)}
\label{alpha}
\end{equation}
where $C$ is an integration constant determined by the initial
condition for the flow at $z=z_h$. For the original theory with
action (\ref{gravityaction}), we impose $\alpha(z_h)=1$ (in
general one can include brane localized curvature terms).
Expanding (\ref{alpha}) in powers of momentum we find
\begin{equation}
\alpha(a)=1+\frac 1 {12} (1-\frac {a^2} {z_h^2})(q a)^2-\frac
{a^4-4 a^2 z_h^2+3z_h^4+4z_h^4 \log(a/z_h)} {96z_h^4} (qa)^4+
{\cal O}((qa)^6).
\end{equation}
This equation is particularly interesting. It tells us that the
hidden brane tension is at a fixed point since it does not scale
with $a$. This is a consequence of the fact that the distance
between the two branes is undetermined in the original theory.

Running the hidden brane to $a \sim z_v$ the bulk contribution to
the action goes to zero. The quadratic action is given by the sum
of the visible and effective brane actions. Up to two derivatives,
the linearized analysis and general covariance completely
determine the effective action. In terms of the field ${\hat
g}_{\mu\nu}(x)=k^2 z_v^2 G_{\mu\nu}(x,z_v)$ we have,
\begin{equation}
S_4=24 \frac {M_5^3} {k^3 z_v^4} \int d^4x \sqrt{-{\hat g}}(-z_v^2
\alpha^{(2)} R({\hat g})-\alpha^{(0)} +1)+ {\cal O} (R({\hat
g})^2) \label{rundownaction}
\end{equation}
Notice that, with the tensions tuned, the visible and effective
hidden brane tensions cancel leaving, as expected, zero
cosmological constant in the four dimensional effective theory.
The value of the $4D$ Planck mass (\ref{planckmass}) can be
readily obtained from the coupling $\alpha^{(2)}(a)$. The
coefficients of the higher derivative operators are not fully
determined by the linearized analysis.

For a general local hidden brane action at $z_h$ equation
(\ref{gravityflow}) dictates a flow for the couplings
$\alpha^{(2j)}(a)$. These couplings scale as $(a/z_h)^{2j}$. The
$q^2$ expansion appears to break down well below the scale of the
lowest lying KK mode $\sim 1/z_v$. As in the case of gauge fields
this effect is related to the shape of the modes in the
extra-dimension: the massless graviton mode is localized near the
hidden brane while the wave function of the KK modes is peaked on
the visible brane.

The graviton zero mode couples to the visible brane a factor of
$z_h/z_v \sim {\cal O}(10^{-15})$ more weakly than the KK modes.
Because of this, at least in principle, in a scattering experiment
with gravitons, deviations from ordinary gravity could be detected
at energies exponentially smaller than the masses of the KK modes.
On the other hand, these KK modes give corrections to the
Newtonian potential only at distances which are parametrically
smaller than the typical mass so they are in practice
unobservable.

To illustrate this explicitly, we consider the exact propagator of
the effective theory. This can be directly obtained from effective
brane coupling evaluated at $z_v$. Suppressing the tensor
structure of the propagator we have:
\begin{equation}
\Delta(q,z_v)= \frac 1 {\alpha-1} = \frac {J_1(q z_h)Y_2(q
z_v)-J_2(q z_v)Y_1(q z_h)} {q z_v(J_1(q z_h)Y_1(q z_v)-J_1(q
z_v)Y_1(q z_h))}
\end{equation}
This propagator has an isolated pole at zero momenta.  The mass
scale of the KK states is set by the visible brane. The propagator
can be rewritten as a sum of contributions from each mode:
\begin{equation}
\Delta(q,z_v)=\sum_n \frac {R(m_n^2)} {q^2-m_n^2}
\end{equation}
where $R(m_n^2)$ is the residue of the propagator at the pole
$m_n^2$. In general, the residue is just (up to normalization) the
square of the coupling of the mode to matter. It is possible to
check that:
\begin{equation}
R(m_n^2)\sim \frac {z_v^2} {z_h^2} R(0)~~~~~~~~n \ge 1
\end{equation}
In the language of modes the interaction lagrangian takes the form
\begin{equation}
{\cal L}=-\frac 1 {M_{4}} h^{(0)}_{\alpha \beta}(x) T^{\alpha
\beta}(x)-\frac 1 {M_{4}} \Big(\frac {z_v} {z_h}\Big)
\sum_{n=1}^\infty h^{(n)}_{\alpha\beta}(x) T^{\alpha\beta}(x)
\end{equation}
reproducing the result in Ref.~\cite{Rizzo2}. In this equation
$T^{\alpha\beta}(x)$ is the energy momentum tensor obtained by
running down the hidden brane so it automatically includes the
effects of bulk matter.


\section{An Improved EFT for Massless Fields}
\label{improvement}

We have seen that the effective field theories derived above are
not valid at energies just below $1/a$, the cutoff expected from holographic
arguments.
 In the case of gravity  the cutoff scale is
exponentially lower. The cutoff is lower by a large $\log$ in the
case of the photon.  In this section we introduce a method of
brane running where the massless zero mode is treated explicitly
as a massless field on the effective hidden brane. This theory is
equivalent to the theory without such a field. However, it
has an attractive fixed point where the massless field decouples
from the bulk field representing the massive KK tower. The cutoff
of the improved theory is  ${\mathcal O}(1/a)$.

\subsection{Introducing the massless brane field}

For simplicity, we introduce this method in the case of a massless
bulk scalar field. Consider a free scalar field  in AdS space.
\begin{eqnarray}
S & = & S_{bulk} + S_{vis} + S_{hid} \nonumber \\
S_{bulk} & = & \frac 1 2 \int d^4 x dz \sqrt{G} \left( G^{MN}
\partial_M
\chi \partial_N \chi \right)  \nonumber \\
S_{hid} & = & -\frac 1 2  \int d^4 x \sqrt{-g_h} \left( \chi k
\lambda_{h2}
\chi \right) \nonumber \\
S_{vis} & = & - \frac 1 2\int d^4 x \sqrt{-g_v} \left( \chi k
\lambda_{v2} \chi \right) \label{scalaraction}
\end{eqnarray}
In the set of effective theories of different radii it was found
in Ref. \cite{runrs} that the quadratic coupling on the effective hidden brane
at $a$ obeys the RG equation
\begin{equation}
a\frac{d \lambda_2} {da} = 4 \lambda_2 - \frac{1}{2} \lambda_2^2
-2 q^2 a^2.
\label{lambdarunning}
\end{equation}

If there is no hidden brane mass term ($\lambda_{h2}^{(0)}=0$) one
will not be generated in the flow.  The theory does not flow to a
fixed point because the fixed point contains a mass term. The
couplings scale as $\lambda_2^{(j)} = (a/z_h)^j$, becoming
exponentially large as we approach the visible brane. This
effective theory suffers from the same problem as the photon and
graviton theories discussed earlier. The cutoff of the effective
field theory is far below the scale $1/a$.

We can find an equivalent description of the theory presented
above by adding a massless hidden brane field. We take for the
hidden brane action
\begin{equation}
S_{hid} = -\frac 1 2 \int d^4 x \sqrt{-g_h} \left(Z_b g^{\mu
\nu}_h
\partial_{\mu} b
\partial_{\nu}b+k (\chi-\sqrt{k}
b) \gamma_h (\chi-\sqrt{k} b)  \right),
\end{equation}
where $b$ is a four dimensional field confined to the hidden
brane, $Z_b$ its field strength and
\begin{equation}
\gamma_h = \gamma_h^{(0)} -k^{-2} \gamma_h^{(2)} g_h^{\mu \nu}
\partial_{\mu} \partial_{\nu} + \ldots
\end{equation}
From the equation of motion for $\chi$ one derives the boundary
condition in momentum space
\begin{equation}
z \partial_z \chi_q(z) \Big|_{z=z_h}=\frac {\gamma_h} 2
(\chi_q(z_h)-\sqrt{k} b_q).
\label{modbound}
\end{equation}
The field $b$ satisfies the equation
\begin{eqnarray}
Z_b q^2 z_h^2 b_q-\gamma(\frac {\chi_q} {\sqrt{k}} -b_q)=0.
\end{eqnarray}
Integrating out $b$ according to its equation of motion and
replacing in (\ref{modbound}) we obtain
\begin{equation}
z \partial_z \chi_q(z) \Big|_{z=z_h}=\frac 1 2 \frac{\gamma_h Z_b
~q^2 a^2}{\gamma_h +Z_b q^2 a^2} \chi_q(z_h).
\end{equation}
This is the same boundary condition as the one implied by the
original action (\ref{scalaraction}) when
\begin{equation}
\lambda_{h2} =\frac{\gamma_h Z_b ~q^2 a^2}{\gamma_h +Z_b q^2 a^2}.
\end{equation}

Now we require in the running of the radius that the effective
theory with the brane field included be equivalent to the
effective theory without that field. This requirement is clearly
\begin{equation}~\label{kappalambda}
\lambda_{2}(a) = \frac{\gamma(a) Z_b(a) q^2a^2}{\gamma(a) + Z_b(a)
q^2 a^2}.
\end{equation}
We would like the field $b$ to represent the massless zero mode.
We therefore require that field strength flows according to
\begin{equation}
a \frac{d Z_b}{da} = 2 Z_{b} - 2.
\end{equation}
i.e. $Z_b$ is identified with the coupling $\lambda_2^{(2)}$ whose
flow can be extracted from (\ref{lambdarunning}). Obeying this
equation, the field strength grows large as $a \to z_v$ and the
field $b$ decouples from $\chi$. The flow for $\gamma$ is
determined by the flow equation for $\lambda_2$ and by equation
(\ref{kappalambda}),
\begin{equation}
a \frac{d \gamma}{da} = 4 \gamma - \frac{1}{2} \gamma^2 - 2 q^2
a^2 - \frac{4 \gamma}{Z_b}.
\end{equation}

The important point in all of this is that the flow equations have
an attractive fixed point given by
\begin{eqnarray}
1/Z_b^* & = & 0, \\
\gamma^* &=& 2 qa \frac{J_{1}(qa)}{J_{2}(qa)} \approx 8-\frac
{(qa)^2} 3-\frac {(qa)^4} {144}+\dots
\end{eqnarray}
Flowing towards this fixed point the massless field $b$ decouples.

The improved effective theory is valid up to the naive scale
$1/a$. In the limit where $a \to z_v$ the field $\tilde{\chi}(x)=
\chi(x,z_v)- \sqrt{k} b(x)$ incorporates the contribution of the
KK modes and $b$ is the massless zero mode.


\subsection{EFT improvement for gauge fields}

We can improve the effective theory running for the $U(1)$ gauge
theory and gravity as we have done in the case of the scalar
field. The basic structure wherein the massless field is added is
the same with additional gauge and index structure. In the case of
the bulk photon we add a four-dimensional $U(1)$ gauge field
localized on the hidden brane. The quadratic hidden brane action
is
\begin{equation}
S_{hid} = -\frac{1}{4 g_5^2 k} \int d^4 x \sqrt{g_h} \left( Z_B
g_h^{\mu \rho} g_h^{\nu \sigma} B_{\mu \nu} B_{\rho \sigma}+ 2k^2
g_h^{\mu \nu} (A_{\mu} - B_{\mu}-\partial_{\mu} \alpha) \rho_h
(A_{\nu} - B_{\nu}-\partial_{\nu} \alpha )
 \right),
\end{equation}
where $B_{\mu}$ is the massless gauge field on the hidden brane
and $B_{\mu \nu} = \partial_{\mu} B_{\nu} - \partial_{\nu}
B_{\mu}$. We have introduced a $4D$ goldstone mode $\alpha$ in
order to break spontaneously one combination of the $U(1)$ gauge
symmetries. In the previous equation
\begin{equation}
\rho_h = \rho_h^{(0)} -k^{-2} \rho_h^{(2)} g_h^{\mu \nu}
\partial_{\mu} \partial_{\nu} + \ldots
\end{equation}
By making an appropriate identification of the couplings of this
theory with the theory of Section \ref{photon}, namely
\begin{equation}
\tau_h = \frac{\rho_h Z_B}{\rho_h + Z_B q^2 z_h^2},
\end{equation}
we can insure that the two are equivalent. Running the radius
down with the requirement
\begin{equation}
\tau(a) = \frac{\rho(a) Z_B(a)}{\rho(a) + Z_B(a) q^2 a^2},
\end{equation}
the couplings flow according to
\begin{eqnarray}
a \frac{d Z_B}{d a} & = &  2 \\
a \frac{d \rho}{da} & = & 2 \rho + \frac{\rho^2}{2} + 2 q^2 a^2 +
4 \frac{\rho}{Z_B},
\end{eqnarray}
where the flow of $Z_B$ is chosen to reproduce the zero mode of
the bulk field. The flow possess an attractive fixed point,
\begin{eqnarray}
1/Z_B^* & = & 0 \\
\rho^* & =& -2 q a \frac{J_0(qa)}{J_1(qa)}\approx -4+\frac
{(qa)^2} 2+\frac {(qa)^4} {48}+\dots
\end{eqnarray}
where the field $B_\mu$ decouples.

We can extend the analysis above to include interactions of the
gauge field on the hidden brane (see the Appendix for a detailed
analysis of interacting gauge theories). We write a hidden brane
action in terms of the fields $\tilde{A}_{\mu} = A_{\mu} -
B_{\mu}-\partial_{\mu} \alpha$ and $B_{\mu \nu}$. We choose the
coefficients of the operators so that this theory is equivalent to
a particular interacting theory upon removing the field $B_{\mu}$.
We know that a fixed point for this theory exists, the quadratic
field fixed point where all coefficients of higher field operators
are zero. Consider now a small perturbation from the fixed point
by a gauge invariant operator schematically of the form $\epsilon
\tilde{A}^n B^m$ where indices and derivatives are suppressed.
Working to first order in $\epsilon$, we remove the field
$B_{\mu}$ and determine the flow for $\epsilon$ from the flow
equation (\ref{photonflow}) presented in the Appendix. The
operator is irrelevant for all $m$ and $n$. Therefore, in the case
of the $U(1)$ gauge theory it is in general possible to find a
sensible low energy effective theory with a cutoff $1/a$.

We can also examine the gravitational theory at the linearized
order by adding a massless brane graviton. As before the effective
theory has an attractive fixed point and the cutoff is $1/a$.

\section{Discussion}
\label{conclusions}

We have extended the RG formalism proposed in \cite{runrs} to
gauge fields and gravity in the RS1 scenario. Moving within the
space of effective field theories with varying radius of the extra
dimension, the hidden brane couplings flow in a way that resembles
a $4D$ RG flow. The flows for the theories we consider do not have
an attractive fixed point due to the presence of a massless zero
mode (photon or graviton). The cutoff of the effective theory with
the hidden brane located at $a$ is lower than the expected scale
$1/a$. We have found an improved procedure where we introduce a
massless hidden brane field. This equivalent theory has an
attractive fixed point. The cutoff is $1/a$. Running the radius
down, the massless field becomes the massless zero mode of the
gauge/gravity theory.

Our method allows one to compute classical effective field
theories for a low energy observer living near the visible brane
in a simple and systematic way. We believe that these techniques
can be used to simplify $5D$ computations. It would be interesting
to extend this formalism to the quantum level. This, for example,
could shed some light on the quantum running of the $4D$ zero mode
gauge coupling, relevant for studies of unification in the RS1
scenario. More general gravitational backgrounds could also be
analyzed using the techniques presented in this paper.

Consideration of the Goldberger-Wise scenario including the
gravitational back reaction and the extension of this work into
the quantum regime has the potential to provide a comprehensive
demonstration of the stability of the hierarchy between the
weak/Planck scales in the RS1 model.

\section*{Acknowledgments}
We would like to thank Raman Sundrum for very useful extended
discussions and for reading the manuscript. We would also like to
thank Kaustubh Agashe for helpful discussions. The research of
A.L. is supported by NSF Grant P420D3620434350. The research of
M.R. is supported by NSF Grant NSF-PHY-9970781 and
NSF-PHY-0099468.

\appendix
\section{Scalar QED}
\label{scalarqed}

In this Appendix we apply the method developed in this paper to
the case of scalar QED. The action is
\begin{equation}
S=S_{bulk} + S_{hid} + S_{vis},
\end{equation}
where
\begin{equation}
S_{bulk}  = \int d^4 x \int_{z_v}^{z_h}dz \sqrt{G}
\left(-\frac{1}{4 g_5^2} G^{MN} G^{PQ} F_{MP} F_{NQ} +  G^{MN}
(D_M \chi) (D_N \chi)^\dagger-m^2 \chi \chi^\dagger \right).
\end{equation}
Here $\chi$ is a massive scalar field charged under the $U(1)$
gauge symmetry. $S_v$ and $S_h$ are the most general brane actions
local in $\chi$, $A_{\mu}$ and their $x$-derivatives and
consistent with gauge invariance. We look for a flow of the hidden
brane couplings that preserves this form. Although the calculation
of this flow is technically more complicated, the procedure is
conceptually the same as in the free photon theory.

The equations of motion
\begin{equation}
\frac{\delta S}{\delta A_{\mu}}=0 \qquad {\rm and} \qquad
\frac{\delta S}{\delta \chi} = 0
\end{equation}
require boundary conditions for $A_{\mu}$ and $\chi$ at the hidden
and visible branes. At the hidden brane the boundary conditions
are
\begin{eqnarray}
z \partial_z A_{\mu} \bigg|_{z=z_h} & = & -\frac{1}{2 \sqrt{-g_h}}
\frac{\delta S_h}{\delta A_{\nu}(x)} g_{\mu \nu}^{h}
\Bigg|_{A_{\nu}(x) = A_{\nu}(x, z_h)}
\\
z \partial \chi \bigg|_{z=z_h} & = & -\frac{1}{2\sqrt{-g_h}}
\frac{\delta S_h}{\delta \chi^*(x)} \Bigg|_{\chi^*(x) =
\chi^*(x,z_h)}.
\end{eqnarray}
Notice that the functional derivatives are with respect to $4D$
fields.

We look for a new local action $S_a$ such that the
boundary conditions
\begin{eqnarray}
z \partial_z A_{\mu}|_{z=a} & = & -\frac{1}{2 \sqrt{-g}}
\frac{\delta S_a}{\delta A_{\nu}(x)}
g_{\mu \nu} \Bigg|_{A_{\nu}(x) = A_{\nu}(x, a)} \\
z \partial \chi|_{z=a} & = & -\frac{1}{2\sqrt{-g}} \frac{\delta
S_a}{\delta \chi^*(x)} \Bigg|_{\chi^*(x) = \chi^*(x,a)}
\end{eqnarray}
are satisfied, where $z_v > a \ge z_h$ and $S_a = S_h$ when
$a=z_h$.  Here $g_{\mu \nu}(x) = G_{\mu \nu}(x,a)$.  The fields
that satisfy these boundary conditions must be solutions to the
original equations of motion in the restricted region $z_v \ge z
\ge a$.

We are interested in the differential flow of $S_a$ as we vary
$a$.  Taking the logarithmic derivative of these equations and
using the equations of motion we write the flow equations in terms
of the brane action. The first boundary condition leads to the
flow equation
\begin{eqnarray}~\label{photonflow}
a \frac{\partial}{\partial a} \left( \frac{\delta S_a}{\delta
A_{\mu}(x)} \right) & = & - 2 \sqrt{-g} g^{\mu \nu} g^{\rho
\sigma} \partial_{\rho} F_{\sigma \nu} + 2 i \sqrt{-g} g^{\mu \nu}
(\chi^* D_\nu \chi - \chi D_v \chi^*)\nonumber \\ && +\int d^4 x'
\Bigg( \frac{\delta^2 S_a}{\delta A_{\mu}(x) \delta \chi(x')}
\left( \frac{1}{2\sqrt{-g}} \frac{\delta S_a}{\delta \chi^*(x')}
\right) + \frac{\delta^2 S_a}{\delta A_{\mu}(x) \delta \chi^*(x')}
\left(
\frac{1}{2\sqrt{-g}} \frac{\delta S_a}{\delta \chi(x')} \right)\nonumber \\
&&- 2 \frac{\delta^2 S_a}{\delta A_{\mu}(x) \delta g^{\rho
\sigma}(x')} g^{\rho \sigma} + \frac{\delta^2 S_a}{\delta
A_{\mu}(x) \delta A_{\rho}(x')} \left( \frac{1}{2 \sqrt{-g}}
\frac{\delta S_a}{\delta A_{\sigma}(x')} \right) g_{\rho \sigma}
\Bigg) .
\end{eqnarray}
where as before $A_{\mu}(x)$ and $\chi(x)$ are $4D$ fields
evaluated such that $A_{\mu}(x) = A_{\mu}(x,a)$ and $\chi(x) =
\chi(x,a)$. From the second boundary condition we obtain
\begin{eqnarray}
a \frac{\partial}{\partial a} \left( \frac{\delta S_a}{\delta
\chi(x)} \right) & = &- 2\sqrt{-g} g^{\mu \nu} D_{\mu} D_{\nu}
\chi^* + 2\sqrt{-g} m^2 \chi^* \nonumber \\ &&+\int d^4 x' \Bigg(
\frac{\delta^2 S_a}{\delta \chi(x) \delta \chi(x')} \left(
\frac{1}{2\sqrt{-g}} \frac{\delta S_a}{\delta \chi^*(x')} \right)
+ \frac{\delta^2 S_a}{\delta \chi(x) \delta \chi^*(x')} \left(
\frac{1}{2\sqrt{-g}} \frac{\delta S_a}{\delta \chi(x')} \right)
\nonumber \\ && - 2 \frac{\delta^2 S_a}{\delta \chi(x) \delta
g^{\rho \sigma}(x')} g^{\rho \sigma} + \frac{\delta^2 S_a}{\delta
A_{\rho}(x') \delta \chi(x)} \left( \frac{1}{2 \sqrt{-g}}
\frac{\delta S_a}{\delta A_{\sigma}(x')} \right) g_{\rho \sigma}
\Bigg) .
\end{eqnarray}
These equations represent the $\beta-$functions for the couplings
of local operators on the effective hidden brane.

Notice that these flow equations are local.  As a consequence, we
will always flow within actions that contain only local operators.
These equations also respect $4D$ gauge invariance,
\begin{eqnarray}
A_{\mu} &\to& A_{\mu} + \partial_{\mu} \Lambda \\
\chi &\to & e^{i \Lambda} \chi
\end{eqnarray}
This implies that if we start with a $4D$ gauge invariant action
at $z_h$ any boundary action consistent with the flow equations
will also be $4D$ gauge invariant. The full action is still $5D$
gauge invariant.

We now show that the effective theory generated is sensible in the
sense that the effects of the higher dimension operators generated
on the effective hidden brane are suppressed at energies below a
cutoff scale of the order $1/(a \log (a/z_h))$. We will do this by
showing that the dimensionless coefficient of no operator flows to
an excessively large number. Therefore one can truncate to the
leading operators as in standard effective field theory.

In the case of the massive free scalar with no photon present,
studied in \cite{runrs}, it was found that there is an attractive
fixed point to which a general local brane action flows. The
coefficient of an operator with $i$ fields and $j$ derivatives
scales as $(a/z_h)^{\gamma_i^{(2j)}}$ where
\begin{equation}
\gamma_i^{(2j)} = 4 - 2j - \frac{i}{2}(4+2 \nu) < 0
\end{equation}
and $\nu=\sqrt{4+m^2/k^2}$. Higher dimension operators are
suppressed for momentum $q$ in the region $qa < 1$.

In the case of a free photon without a scalar field we have found
by explicit calculation in Section \ref{photon} that the
coefficient of an operator with $2k$ derivatives scales as $(\log
(a/z_h))^k$. If we restrict momentum such that $q^2 a^2 \log
(a/z_h) < 1$, higher derivative operators will be unimportant. The
cutoff here is parametrically smaller than $1/a$.

Now consider operators with $j$ derivatives and $i$ powers of the
$A_{\mu}$ field.  Let the coefficient of such an operator be
denoted $\tau_{i,j}$.  The flow of the coefficients of these
operators will be dictated by equation (\ref{photonflow}). In the
absence of scalars this gives a flow of the form
\begin{equation}
a \partial_a \tau_{i, j} = -j~ \tau_{i, j} + \sum_{klmn} c_{klmn}
\tau_{k, l} \tau_{m, n}
\end{equation}
where $c_{klmn}$ are order one numbers. We must have $k+m=i+2$,
$n+l=j$, $ i \le j$, $k \le l$ and $m \le n$.  These conditions
are sufficient to inductively show that $\tau_{i,j}$ scales as
$(a/z_h)^{-i} (\log (a/z_h))^{(j-i)/2}$ for $i \ge 4$. These
operators are suppressed as we move the effective hidden brane
closer to the visible brane.

We now consider those brane operators with interaction terms
between the photon and scalar.  Some of these operators will be
contained in the gauge invariant operators with covariant
derivatives of $\chi$ only. Coefficients of these operators must
flow as in the case of the self-interacting scalar without the
photon field by gauge invariance.  Operators not of this form (for
example $|\chi|^2 F^2$) must be examined by studying equation
(\ref{photonflow}) to determine the flows of their coefficients.
We denote the coefficient of an operator with $i$ derivatives, $j$
powers of $A^{\mu}$ and $k$ powers of $\chi$ as $\alpha_{i,j,k}$.
The flow equation for these coefficients takes the form
\begin{equation}
a \partial_a \alpha_{i,j,k} = \gamma_{ijk} \alpha_{i,j,k} + \delta
\lambda_2^{(0)} \alpha_{i,j,k} + \sum c_{lmnpqr} \alpha_{l,m,n}
\alpha_{p,q,r}
\end{equation}
where
\begin{equation}
\gamma_{ijk}=4-(i+j)-k(2+\nu) < 0
\end{equation}
Here the indices are constrained to obey $l+p = i$, $m+q=j+2$,
$n+r=k$, $i \ge j$, $l \ge m$, $p \ge q$. By induction the terms
with the strongest $a$ dependence scale as
 $\alpha_{ijk} \sim (a/z_h)^{4-k(2+\nu)-j} (\log (a/z_h))^{(i-j)/2}$
for $k \ge 2$.

The important thing to take away from this discussion is that the
effective theory with hidden brane located at $a$ will be valid up
to momenta satisfying $q^2 a^2 \log (a/z_h) < 1$. The higher
derivative operators will be suppressed at low energy allowing for
simplification in effective field theory computations.

\end{document}